\documentclass[aps,prb,twocolumn,showpacs,superscriptaddress,amsmath,amssymb]{revtex4-1}
\usepackage[pdftex]{graphicx}
\usepackage[pdftex,bookmarks=true,bookmarksopen,bookmarksnumbered,
                colorlinks,
                linkcolor=blue,
                citecolor=blue]{hyperref}
\usepackage{dcolumn}
\usepackage{bm}

\begin{document}


\title{Robust two-qubit gates for donors in silicon controlled by hyperfine interactions}

\author{Rachpon Kalra}
\affiliation{Centre for Quantum Computation and Communication Technology, School of Electrical Engineering \& Telecommunications, University of New South Wales, Sydney, New South Wales 2052, Australia}
\author{Arne Laucht}
\affiliation{Centre for Quantum Computation and Communication Technology, School of Electrical Engineering \& Telecommunications, University of New South Wales, Sydney, New South Wales 2052, Australia}
\author{Charles Hill}
\affiliation{Centre for Quantum Computation and Communication Technology, School of Physics, University of Melbourne,
Melbourne, Victoria 3010, Australia}
\author{Andrea Morello}
\email{a.morello@unsw.edu.au}
\affiliation{Centre for Quantum Computation and Communication Technology, School of Electrical Engineering \& Telecommunications, University of New South Wales, Sydney, New South Wales 2052, Australia}

\date{\today}

\begin{abstract}
We present two strategies for performing two-qubit operations on the electron spins of an exchange-coupled pair of donors in silicon, using the ability to set the donor nuclear spins in arbitrary states. The effective magnetic detuning of the two electron qubits is provided by the hyperfine interaction when the two nuclei are prepared in opposite spin states. This can be exploited to switch on and off SWAP operations with modest tuning of the electron exchange interaction. Furthermore, the hyperfine detuning enables high-fidelity conditional rotation gates based on selective resonant excitation. The latter requires no dynamic tuning of the exchange interaction at all, and offers a very attractive scheme to implement two-qubit logic gates under realistic experimental conditions.
\end{abstract}

\maketitle
\section{Introduction}\label{intro}
The electron spin of a donor atom in silicon represents a natural, highly-coherent quantum bit, bound to a well-defined confining potential and hosted by the most important material in modern technology. The recent demonstrations of high-fidelity single-shot readout \cite{Morello2010} and control of both the electron \cite{Pla2012} and the nuclear \cite{Pla2013} spins of a $^{31}$P donor in a silicon nanostructure have added tremendous momentum to this quantum computer architecture \cite{Kane1998, Awschalom2013}.  The next step towards constructing a universal set of quantum gates is to demonstrate two-qubit logic operations \cite{DiVincenzo1995}.  This has been accomplished in several architectures including those of photonic qubits \cite{OBrien2003}, superconducting circuits \cite{Plantenberg2007}, qubits defined in quantum dots \cite{Nowack2011,Shulman2012}, atoms in electromagnetic traps \cite{Monroe2013}, and nitrogen-vacancy centres in diamond \cite{Neumann2010}.  While several proposals for the implementation of a two-qubit gate with donor electrons exist \cite{Hill2005,Hollenberg2006,Hill2007}, they pose very challenging demands on the tunability of the spin exchange interaction $J$, assumed to be switchable from $\sim 0$ to  $> 1$~GHz. It is also predicted that $J$ can vary strongly upon displacing a donor by even a single lattice site \cite{Koiller2002, Wellard2003PRB}, thus requiring true atomic precision in the placement of the donors. These considerations have contributed to create some skepticism on the viability of donor spin qubits for large quantum computer architectures.

Here we propose two implementations of two-qubit gates that overcome these challenges.  Both these gates, when combined with previously demonstrated single qubit operations \cite{Pla2012}, are universal for quantum computing. Our proposals are based on exploiting the hyperfine interaction with the donor nuclear spins, and the ability to control the nuclear spin state.  High-fidelity control and readout of a single $^{31}$P nuclear spin has been established experimentally \cite{Pla2013}, validating our main assumption. It was also found that a nuclear spin prepared in an eigenstate maintains it state for several minutes \cite{Pla2013}. The core of the idea is to prepare the nuclei in opposite spin states, so that the hyperfine coupling provides a substantial difference in the local magnetic field experienced by the electrons.  Magnetically detuning the energies of electron spin qubits has been proposed \cite{Meunier2011} and implemented in several ways, including the fabrication of a micromagnet adjacent to the qubits \cite{Pioro2008}, introducing an inequivalence in $g$-factors \cite{Luis2011}, or dynamically polarizing the background nuclear spin bath \cite{Foletti2009}.  In comparison, the initialization of the two nuclei in the two-donor system presents an extremely compact, consistent and easily switchable source of magnetic detuning. In the first proposal we focus on using the hyperfine interaction to switch the amplitude of exchange oscillations to perform a $\sqrt{\text{SWAP}}$ gate.  This requires a reasonable two orders of magnitude control of $J$, which could be achieved with an easily fabricable device design.  The second two-donor gate is a prototypical implementation of a conditional rotation (CROT), as demonstrated for superconducting qubits \cite{Plantenberg2007} and spin-qubits in diamond \cite{Jelezko2004}.  This is the resonant rotation of one qubit conditional upon the state of the other.  We show that high-fidelity entangling two-qubit gates can be performed between donor pairs. Dynamic control of the exchange coupling is not required at all in this case, and high-fidelity gates can be achieved for a wide range of coupling strengths. These gates can tolerate over two orders of magnitude of variability in $J$, which means that atomically-precise donor placement is not required.  The two-qubit gates described here can be applied locally, in separate interaction regions. Spin transport, possibly via CTAP rails \cite{Greentree2004} or spin-buses \cite{Friesen2007}, as in the framework proposed by Hollenberg \textit{et al.} \cite{Hollenberg2006} could be utilized to implement a scalable quantum computing architecture.

\section{Proposed system and theoretical representation}
We consider a system of two donors described by the following Hamiltonian (in units of frequency):
\begin{eqnarray}\nonumber
H &=& \gamma_e B_0\left(S_{1_z}+S_{2_z}\right)+\gamma_nB_0\left(I_{1_z}+I_{2_z}\right)\\
&&+A_1\left(\mathbf{S}_1\cdot\mathbf{I}_1\right)+A_2\left(\mathbf{S}_2\cdot\mathbf{I}_2\right)+J\left(\mathbf{S}_1\cdot\mathbf{S}_2\right),
\label{Hami}
\end{eqnarray}
where subscript 1 (or 2) refers to donor-1 (or 2), $\gamma_e$ (or $\gamma_n$) is the gyromagnetic ratio of the electrons (or nuclei), $B_0$ is the externally applied magnetic field, $S$ (or $I$) is the electron (or nuclear) spin operator with $z$-component $S_z$ (or $I_z$), and $A$ is the hyperfine interaction. In the following we will focus on a pair of $^{31}\text{P}$ donors in silicon, but our conclusions are also valid for other donor species in silicon. We allow for different values of $A$ in the two donors, since the different local electric field can Stark-shift the hyperfine coupling \cite{Rahman2007PRL}. We define $\Delta{A} = (A_2 - A_1)/2$ and $\bar{A} = (A_1 + A_2)/2$.  The bulk value for $A$ for $^{31}\text{P}$ donors is 117 MHz and we assume $|\Delta{A}|/\bar{A}$ to be in the range of 1-4\%, i.e. $\Delta{A} \sim$~a few MHz, as expected for $^{31}\text{P}$ donors spaced by $\sim 20$~nm in a similar nanostructure \cite{Mohiyaddin2013}. The parameters $J, A_1, A_2$ depend on local electric fields and the exact positions of the two donors, and can be extracted from an experiment which we describe in Sec.~\ref{fingerprint}.

The coupled donor-pair spin Hamiltonian has been studied by Refs. [\onlinecite{Slichter1955,Jerome1964,Marko1968,Shimizu1968,Pifer1985}] in a high magnetic field such that $\gamma_eB_0\gg J$ and $\bar{A}$. At high fields, where $(\gamma_e - \gamma_n)B_0\gg \bar{A}$, the electrons and nuclei are sufficiently detuned from hyperfine mixing such that their eigenstates can mostly be treated separately.  The effect of the relative strengths of $J$ and $\bar{A}$ on the dynamics of the electrons is the foundation of the proposals of this paper.
The chosen qubit is the spin state of the donor electron, with basis states $|{\uparrow\rangle}$ and $|{\downarrow\rangle}$.  The computational basis states for two-qubit space are thus $|{\uparrow\uparrow\rangle}$, $|{\uparrow\downarrow\rangle}$, $|{\downarrow\uparrow\rangle}$ and $|{\downarrow\downarrow\rangle}$.

\section{Hyperfine-regulated \textit{SWAP} gates}
\begin{figure}[t!]
\includegraphics[width=1\columnwidth]{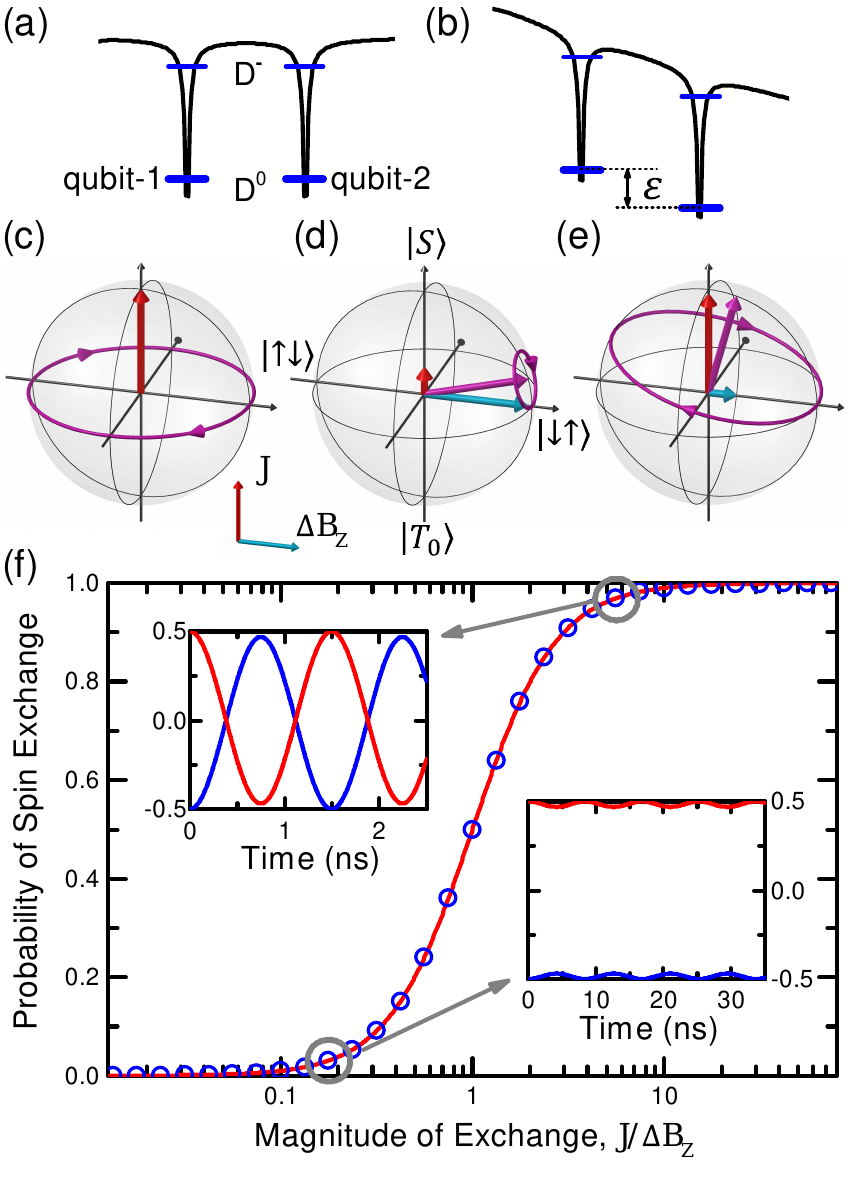}
\caption{\label{figure2} (a),(b) Schematic of the conduction band profile of the coupled donor pair with qubit-1 and qubit-2 (a) in resonance and (b) detuned by $\varepsilon$. (c)-(e) Precession on the $S$-$T_0$ Bloch spheres with initial state $|{\downarrow\uparrow\rangle}$ for exchange-coupled electrons (c) in the absence of coupling to nuclei, (d) where $J < \Delta B_z$ and (e) where $J > \Delta B_z$.  (f) Maximum probability of spin-state exchange between the two donor electrons as a function of $J$ normalized to $\Delta B_z$.  The insets show the time evolution of the expectation value of the electrons initialized as $|{\downarrow\uparrow\rangle}$ for two values of $J$, illustrating the ability to control the amplitude of exchange oscillations.}
\end{figure}

Previous donor-based qubit proposals suggested tuning $J$ via direct modification of the tunnel-barrier \cite{Kane1998,Wellard2003,Hollenberg2006}, requiring precise placement of a $J$-gate between the coupled donors.  Instead we suggest tuning $J$ by detuning the donor potentials by an amount $\varepsilon = E^0_1 - E^0_2$, where $E^0$ is the electrochemical potential of each donor in the neutral $D^0$ state [see Fig.~\ref{figure2}(a-b)]. This method is widely used in double quantum dot systems in the singlet-triplet configuration \cite{Petta2005}, however there the `detuning' is defined as the energy difference between the (1,1) and (0,2) charge configurations. Since in our scheme each donor spin represents a qubit, we do not advocate coming too close to the (0,2) charge configuration -- this would correspond to moving the donor pair from a (D$^0$,D$^0$) to a (D$^+$,D$^-$) state. Nevertheless $J(\varepsilon)$ can be significantly tuned \cite{Burkard1999} from its minimum value at $\varepsilon \approx 0$, to a higher value at $\varepsilon < E_c$, where $E_c \sim 35$~meV is the donor charging energy \cite{Tan2010}. This significantly relaxes the requirements on the nanofabrication, since the control gates only need to be adjacent to the donor pair.

 In the first proposal of this paper, we show how SWAP operations can be switched on and off with modest control of $J(\varepsilon)$.  A perfect SWAP$^\alpha$ operation is a rotation of angle $\alpha\pi$ exactly about the $J$-axis in the $S$-$T_0$ Bloch sphere [see Fig.~\ref{figure2}(c)].  Existing proposals rely on gate-control of $J$ to vary the frequency of the exchange oscillations from (ideally) zero to a maximum value $J_{\text{on}}$. Any residual interaction $J_{\text{off}}$ after the SWAP$^\alpha$ operation would result in further -- unwanted -- evolution of the qubits.  For example, the qubit readout method based on spin-dependent tunneling requires a wait time of order 10~$\mu$s -- 1~ms between the end of the operation and the readout event \cite{Morello2010}. Performing a $\sqrt{\mathrm{SWAP}}$ operation in 10 ns requires $J_{\text{on}} = 25$~MHz, but ensuring that the resulting states have not changed by more than 1\% after 1 ms requires $J_{\text{off}} < 32$~Hz: a six order of magnitude dynamic range which is extremely challenging to achieve.

 Our proposal focuses on controlling the amplitude of the exchange oscillations instead.  We first note that the nuclear spins are also subject to a mutual coupling $J_n$ mediated by $A$ and $J$ \cite{Kane1998}. However this coupling is very small -- a few kHz even for $J \gg 1$~GHz, so we can assume $J_n$ to be negligible relative to the other energy terms of the nuclear states. Recalling that the nuclear states have negligible mixing with the electronic states in a high magnetic field, we can say that the nuclear eigenstates are, with very good approximation, the separable $|{\Uparrow\Uparrow\rangle}$, $|{\Uparrow\Downarrow\rangle}$, $|{\Downarrow\Uparrow\rangle}$ and $|{\Downarrow\Downarrow\rangle}$ states.  The nuclei can be initialized in any of their eigenstates by ionizing the donors one at a time and using the techniques recently demonstrated for the readout and control of the nucleus of a single ionized donor in this architecture \cite{Pla2013}.  Once initialized, the nuclei can be treated as static, contributing to a magnetic field difference for the two electrons, $\Delta B_z = |{\langle\uparrow\downarrow|H|\uparrow\downarrow\rangle -  \langle\downarrow\uparrow|H|\downarrow\uparrow\rangle}|$.  When the nuclei are parallel ($|{\Uparrow\Uparrow\rangle}$ or $|{\Downarrow\Downarrow\rangle}$), $\Delta B_z = \Delta A$, and when they are antiparallel ($|{\Uparrow\Downarrow\rangle}$ or $|{\Downarrow\Uparrow\rangle}$), $\Delta B_z = \bar{A}$.  We can, therefore, switch the strength of $\Delta B_z$ by $1-2$ orders of magnitude -- in a `digital' fashion -- through preparation of the nuclear states.

Tuning $J$ such that $J \ll \Delta B_z$ blocks the exchange oscillations [see Fig.~\ref{figure2}(d)].  To achieve SWAP$^\alpha$ operations, we pulse $J$ so that it is larger than $\Delta B_z$.  The two-electron state here precesses about an axis that tends towards the $J$-axis for increasing $J/\Delta B_z$, as shown in Fig.~\ref{figure2}(e), thus performing exchange oscillations.  After a time $\tau = \alpha/(2\sqrt{J^2+\Delta B_z ^2})\approx \alpha/(2J)$, the system is pulsed back to the low $J$ regime, completing the SWAP$^\alpha$ operation.  Fig.~\ref{figure2}(f) shows calculations and simulations of the fidelity in the `on' and `off' regimes.  The probability of spin exchange is plotted on the vertical axis and $J/\Delta B_z$ on the horizontal axis, where $\Delta B_z$ is $\Delta A$ for parallel nuclei and $\bar{A}$ for anti-parallel nuclei.  The insets of Fig.~\ref{figure2}(f) show the evolution of the expectation value of the two electrons, where $\langle S_z \rangle$ is plotted for electron 1 (blue line) and 2 (red line) initialized in the $|{\downarrow\uparrow\rangle}$ state.  The blue circles in Fig.~\ref{figure2}(f) correspond to the amplitude of the oscillations in these time-evolution simulations.  The Rabi formula for the $S$-$T_0$ Bloch sphere, $J^2/(J^2 + \Delta B_z ^2)$ follows these circles very closely, validating our simplified picture of the system.  The results in Fig.~\ref{figure2}(f) show that SWAP operations can be switched on and off with a fidelity of 99\% by pulsing $J$ between $\Delta B_z/10$ and $10\Delta B_z$ -- two orders of magnitude control of $J$ is sufficient. An alternative method to perform SWAP$^\alpha$ operations would be to use Euler angle construction \cite{Hanson2007}, allowing for exact rotations about the $J$-axis and further reducing the tuning capabilities required.


\section{\textit{CROT} gates}
The CROT operation is another two-qubit gate that can be achieved with the exchange-coupled two-donor system.  Importantly, our proposed realization of the operation does not require any tuning of $J$, further simplifying its practical implementation.  In demonstrating how the system can be used to implement a CROT gate, we make some necessary approximations and quantify the associated errors.

\begin{figure}[t!]
\includegraphics[width=1\columnwidth]{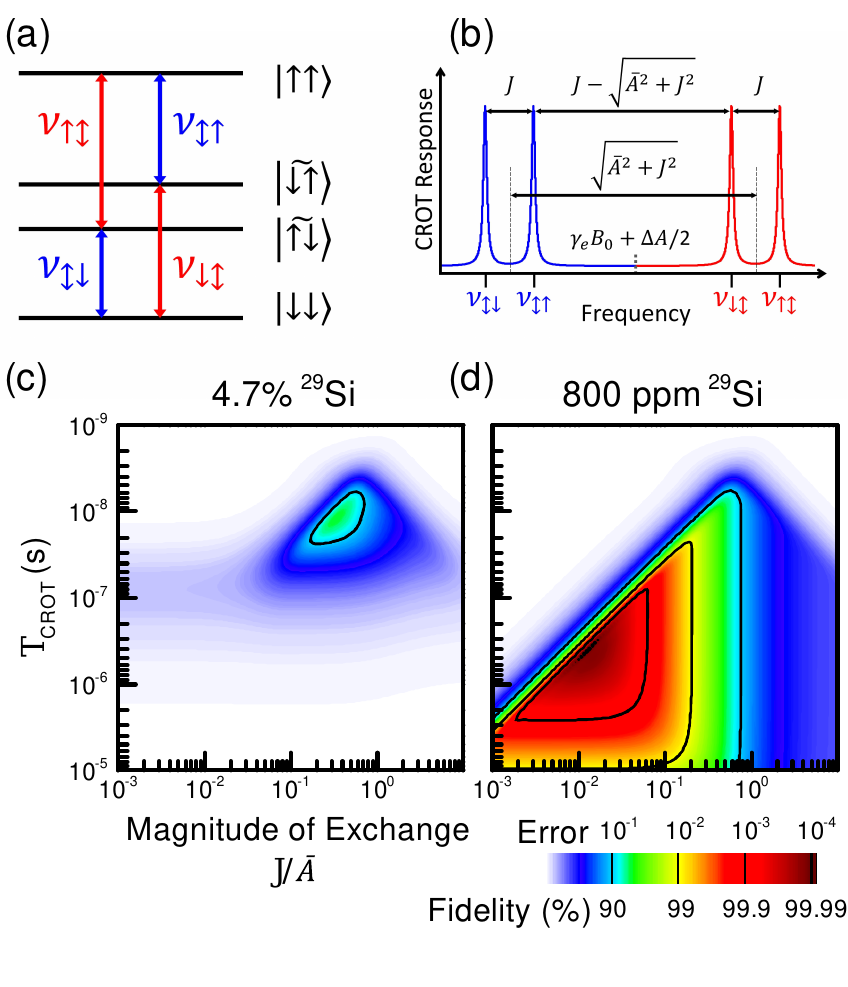}
\caption{\label{figure3} (a) Schematic of the level diagram for the two-donor system with the nuclei prepared in the $|{\Downarrow\Uparrow\rangle}$ state.  (b) Schematic of the corresponding ESR spectrum.  (c),(d) Contour plots of the fidelity of the proposed CROT gate as a function of $J$ and $T_\text{\tiny{CROT}}$, calculated on the basis of the experimental values of the ESR linewidths in (c) natural silicon and (d) isotopically-purified silicon.}
\end{figure}

Fidelity is most often calculated as average gate fidelity over all input states~\cite{Vandersypen2005}.  Perhaps a more meaningful quantity is the minimum fidelity considering all possible input states, though this is more difficult to calculate.  We choose to provide an intuitive measure of the operator fidelity that approximates the minimum fidelity.  Our method is to calculate the total fidelity $F$ by adding, as independent events, the worst-case errors associated with each approximation we make to the Hamiltonian of the system.  The worst-case fidelity for each approximation is defined as
$\min_{\psi_i}(|{\langle \psi_{actual} | \psi_{ideal} \rangle}|^2)$, where $\psi_{ideal}$ and $\psi_{actual}$ are the output states of the operator with and without the approximation, given an input state $\psi_i$.  The input state that yields the minimum fidelity is easy to identify for each individual approximation.  The total fidelity $F$ is then plotted in Fig.~\ref{figure3}(c-d) as a function of $J$ and $T_\text{\tiny{CROT}}$, the gate operation time. The two panels show the fidelities for two concentrations of $^{29}$Si, as will be explained later.

We propose to operate the CROT gate under the condition $J < \bar{A} \ll \gamma_{e}B_0$.  We prepare the nuclei in either the  $|{\Uparrow\Downarrow\rangle}$ or $|{\Downarrow\Uparrow\rangle}$ eigenstate, where they are static and do not participate in the dynamics of the electrons.  The $\Delta B_z$ experienced by the two electrons for this nuclear configuration is $\bar{A}$.  We thus define an electron-only Hamiltonian in the computational basis with the nuclei initialized in the $|{\Downarrow\Uparrow\rangle}$ state,
\begin{equation}
\mbox{$H_1$}  = \left[
\begin{array}{cccc}
E_{\uparrow\uparrow} & 0  & 0 & 0 \\
0 & E_{\uparrow\downarrow} & J/2 & 0\\
0 & J/2 & E_{\downarrow\uparrow} & 0\\
0 & 0 & 0 & E_{\downarrow\downarrow}
\end{array}
\right],
\label{eq:Hamiltonian_General}
\end{equation}
where $E_{\uparrow\uparrow} = \gamma_e B_0+\frac{J}{4}+\frac{\Delta A}{2}$, $E_{\uparrow\downarrow} = \frac{-J}{4}+\frac{-\bar{A}}{2}$, $E_{\downarrow\uparrow} = \frac{-J}{4}+\frac{\bar{A}}{2}$ and $E_{\downarrow\downarrow} = -\gamma_e B_0+\frac{J}{4}+\frac{-\Delta A}{2}$.  We rotate away the $J$-terms leaving the Hamiltonian diagonalized using the change of basis matrix with the eigenstates of the Hamiltonian: $|{\uparrow\uparrow\rangle}$, $|{\downarrow\downarrow\rangle}$ and
\begin{eqnarray}
\label{eq:JawayStates}
|{\widetilde{\uparrow\downarrow}\rangle} = \cos\theta|{{\uparrow\downarrow}\rangle} - \sin\theta|{{\downarrow\uparrow}\rangle}, \\
|{\widetilde{\downarrow\uparrow}\rangle} = \cos\theta|{{\downarrow\uparrow}\rangle} + \sin\theta|{{\uparrow\downarrow}\rangle},
\end{eqnarray}
where $\tan 2\theta = J/\bar{A}$.  The corresponding eigenenergies are $E_{\widetilde{\uparrow\downarrow}} = -J/4 - \sqrt{\bar{A}^2+J^2}/2$ and $E_{\widetilde{\downarrow\uparrow}} = -J/4 + \sqrt{\bar{A}^2+J^2}/2$.

Fig.~\ref{figure3}(a) shows a level diagram of the four eigenstates of the system including the allowed electron spin resonance (ESR) transitions, and Fig.~\ref{figure3}(b) shows a schematic of the corresponding ESR spectrum.  We define the notation e.g. $\nu_{\uparrow\updownarrow}$ to be the transition frequency corresponding to rotating the second electron when the first electron is $|{\uparrow\rangle}$.  We see that
\begin{eqnarray}
\label{eq:NusInRotating}
\nu_{\uparrow\updownarrow} = \gamma_e B_0 + \frac{\Delta A}{2} + \frac{J}{2} + \frac{\sqrt{\bar{A}^2+J^2}}{2},\\
\nu_{\downarrow\updownarrow} = \gamma_e B_0 + \frac{\Delta A}{2} - \frac{J}{2} + \frac{\sqrt{\bar{A}^2+J^2}}{2},\\
\nu_{\updownarrow\uparrow} = \gamma_e B_0 + \frac{\Delta A}{2} + \frac{J}{2} - \frac{\sqrt{\bar{A}^2+J^2}}{2},\\
\nu_{\updownarrow\downarrow} = \gamma_e B_0 + \frac{\Delta A}{2} - \frac{J}{2} - \frac{\sqrt{\bar{A}^2+J^2}}{2}.
\end{eqnarray}
We define our CROT operation to be a $\pi$-rotation at $\nu_{\downarrow\updownarrow}$. An ideal CROT, however, would be a $\pi$-rotation between $|{\downarrow\downarrow\rangle}$ and $|{{\downarrow\uparrow}\rangle}$, not between $|{\downarrow\downarrow\rangle}$ and $|{\widetilde{\downarrow\uparrow}\rangle}$. Using the protocol outlined earlier for calculating the associated error, clearly $\psi_{actual}$ and $\psi_{ideal}$ are $|\widetilde{\downarrow\uparrow}\rangle$ and $|{\downarrow\uparrow}\rangle$.  The inherent error introduced in the operation is $\sin^2 \theta$.  This error increases with $J$ and is independent of $T_\text{\tiny{CROT}}$ and the material, thus appearing as a vertical boundary in the high-$J$ region of the fidelity plots,  Fig.~\ref{figure3}(c-d).

The CROT operation is obtained by applying a magnetic field, rotating at frequency $\nu$ in the plane perpendicular to the $\mathbf{B}_0$, with time-dependent amplitude $B_1(t)$. Transforming this perturbation into a dressed basis, the Hamiltonian in the rotating frame is:
\begin{equation}
\mbox{$H_2$} =\left[
\begin{array}{cccc}
E_{\uparrow\uparrow}-\nu & \gamma_e B_1(t)\mu_S  & \gamma_e B_1(t)\mu_T & 0 \\
\gamma_e B_1(t)\mu_S & E_{\widetilde{\uparrow\downarrow}} & 0 & \gamma_e B_1(t)\mu_S \\
\gamma_e B_1(t)\mu_T & 0 & E_{\widetilde{\downarrow\uparrow}} & \gamma_e B_1(t)\mu_T \\
0 & \gamma_e B_1(t)\mu_S & \gamma_e B_1(t)\mu_T & E_{\downarrow\downarrow}+\nu
\end{array}
\right],
\label{eq:Hamiltonian_RotFrameJaway}
\end{equation}

where $\mu_S = (\cos\theta-\sin\theta)/2$ and $\mu_T =(\cos\theta+\sin\theta)/2$.  We now have $\gamma_e B_1(t)$ coupling the four transitions $\nu_{\uparrow\updownarrow}$, $\nu_{\downarrow\updownarrow}$, $\nu_{\updownarrow\uparrow}$ and $\nu_{\updownarrow\downarrow}$, with different apparent amplitudes $\left[\gamma_e B_1(t)(\cos\theta \pm \sin\theta)/2\right]$ depending on the states involved.  Our next approximations will be to remove these coupling terms in the Hamiltonian for all off-resonant transitions.  To quantify the error associated with these approximations, we treat each transition as an independent qubit system with relevant coupling and detuning, and determine the transition probability given $\gamma_e B_1 (t)$.  Non-zero probabilities at $\nu_{\uparrow\updownarrow}$, $\nu_{\updownarrow\uparrow}$ and $\nu_{\updownarrow\downarrow}$ are considered errors.  In addition to this, a non-unity probability at $\nu_{\downarrow\updownarrow}$ (the chosen CROT frequency) is also an error-event.

We assume that the CROT gate at frequency $\nu_{\downarrow\updownarrow}$ is obtained by applying a resonant microwave pulse with a Gaussian envelope. Its excitation profile, $f_V$ \cite{Vasilev2004}, is many orders of magnitude more selective than that of a square pulse, especially at large detunings.  Other shapes with similar selectivity, such as the Hermite pulse, are also possible candidates \cite{Vandersypen2005} for the CROT gate.  The excitation amplitude we choose is defined as:
\begin{eqnarray}
\label{eq:GaussExc}
B_1(t) = B_1^\text{\tiny{max}} \text{exp}\left( \frac{(t - T_\text{\tiny{CROT}}/2)^2}{2(T_\text{\tiny{CROT}} /6)^2} \right), t \in [0,T_\text{\tiny{CROT}}].
\end{eqnarray}
$T_\text{\tiny{CROT}}$ is clipped at six times the standard deviation of the Gaussian to sufficiently approximate the function.  The time $T_\text{\tiny{CROT}}$ necessary for a $\pi$-rotation is inversely related to $B_1^\text{\tiny{max}}$ -- the complete relationship is given in Ref.~[\onlinecite{Vasilev2004}]. For example, to perform a $\pi$-rotation in 100 ns, $B_1^\text{\tiny{max}}$ would need to be 0.43 mT for a Gaussian pulse and 0.18 mT for a square pulse.  We also take into account the modification of the apparent amplitude of $B_1$ depending on the states involved, $\mu(\theta)$.  We define the probability that a $\pi$-rotation occurs for a certain transition, with frequency $\nu_i$ and corresponding $\mu(\theta)$, as:
\begin{eqnarray}
\label{eq:ExcProb}
p_\pi=\int_{-\infty}^\infty {f_V{(\nu_{\downarrow\updownarrow},T_{\text{\tiny{CROT}}}, \mu(\theta),\nu)}P{(\nu_i, \sigma,\nu)}d\nu}.
\end{eqnarray}
Here, $P{(\nu_i, \sigma,\nu)}d\nu$ describes the broadening of the resonance, a normalized Gaussian distribution with standard deviation $\sigma$, centered at $\nu_i$ \cite{Pla2012}.  The first source of broadening is fluctuations in the surrounding bath of $^{29}$Si spins \cite{Abe2010}.  As the contribution to the broadening depends on the $^{29}$Si concentration, we calculate $F$ for qubits in natural silicon ($^{\text{nat}}$Si), which contains 4.7\% spin-carrying $^{29}$Si nuclei [see Fig.~\ref{figure3}(c)], and in isotopically purified $^{28}$Si with 800 ppm residual $^{29}$Si atoms ($^\text{iso}$Si) [see Fig.~\ref{figure3}(d)].  The other possible source of broadening is the modulation of $A$, $\gamma_e$ or $J$ due to coupling to electric field noise.  We use the linewidths obtained from ESR data on single $^{31}$P donors in gated silicon nanostructures, $\sigma=3.2$ MHz  for $^\text{nat}$Si \cite{Pla2012} and $\sigma = 2$ kHz for $^\text{iso}$Si~\cite{Muhonen2013}.  These experimental values inherently include broadening due to the spin bath, and the modulation of $A$ and $\gamma_e$ via electric field noise, but not of $J$.  The effect on $J$ is difficult to predict, and we neglect it here.

Looking at the fidelity plots, we can see the error associated with the partial excitation (non-zero $p_\pi$) of off-resonant transitions.  The diagonal fidelity-boundary at the top-left is due to the proximity of  $\nu_{\uparrow\updownarrow}$ to $\nu_{\downarrow\updownarrow}$.  Faster gates, corresponding to shorter $T_\text{\tiny{CROT}}$ and broader excitation profiles, require a higher $J$ to separate the two resonances.  The diagonal fidelity-boundary at the top-right is due to the $\nu_{\updownarrow\uparrow}$ coming closer to $\nu_{\downarrow\updownarrow}$ as $J$ increases.  These two boundaries appear at the same positions for both $^{\text{nat}}$Si and $^{\text{iso}}$Si, since they mainly depend on the spectral separation of the lines given by $J$ and $\bar{A}$ which, in both cases, is larger than the quoted linewidths for the inhomogeneous broadening.

The final type of error is the incomplete excitation (non-unity $p_\pi$) of the CROT transition.
The excitation profile must be sufficiently wide (short $T_\text{\tiny{CROT}}$) as compared to the  inhomogeneous broadening to successfully drive the $\pi$-rotation at $\nu_{\downarrow\updownarrow}$.  This results in the horizontal fidelity-boundary at the bottom of Fig.~\ref{figure3}(c-d).  We see that this boundary allows for longer $T_\text{\tiny{CROT}}$ in $^{\text{iso}}$Si [Fig.~\ref{figure3}(d)] as compared to $^{\text{nat}}$Si [Fig.~\ref{figure3}(c)], effectively `unveiling' a large region of high fidelities.  In natural silicon, fidelities of $\sim$95\% are achievable for a range of $J$ values over almost an order of magnitude, with gate times around 30~ns.  In the isotopically-purified material, the peak fidelity achievable exceeds 99.99\% with a gate time of 400~ns.  Fidelities greater than $99.9\%$ are achievable for a range of $J$ values over $\sim$1.5 orders of magnitude, with gate times as short as 80 ns. From a practical perspective, the important result is that fidelities greater than $99\%$ are achievable for a range of $J$ values varying over $\sim$2.6 orders of magnitude. This means that even fabrication methods such as ion implantation \cite{Jamieson2005,Persaud2005,Alves2013}, which inherently suffer from imprecisions in the donor placement, become realistically suitable for multi-qubit donor structures. Also, a recent proposal shows that donor pairs can be exchange-coupled via an intermediate multi-electron quantum dot \cite{Srinivasa2013}. The typical coupling strengths $J \sim 100$~kHz in that proposal would yield $\sim 99.9$\% fidelity for the CROT gates in $^{\text{iso}}$Si described here. Given the demonstrated ability to fabricate top-gated few-electron quantum dots in silicon \cite{Lim2009} that can be tunnel-coupled to donors \cite{Morello2010}, donor-dot hybrids could constitute an appealing new strategy for fabrication and scale-up.

Having quantified the errors associated with applying the excitation, we can make the secular approximation to the Hamiltonian with $\nu  = \nu_{\downarrow\updownarrow}= E_{\widetilde{\downarrow\uparrow}} - E_{\downarrow\downarrow}$,
\begin{equation}
\mbox{$H_3$}  = \left[
\begin{array}{cccc}
E_{\uparrow\uparrow}-(E_{\widetilde{\downarrow\uparrow}} - E_{\downarrow\downarrow}) & 0  & 0 & 0 \\
0 & E_{\widetilde{\uparrow\downarrow}} & 0 & 0 \\
0 & 0 & E_{\widetilde{\downarrow\uparrow}} & \gamma_e B_1(t)\mu_T \\
0 & 0 & \gamma_e B_1(t)\mu_T & E_{\widetilde{\downarrow\uparrow}}
\end{array}
\right].
\label{eq:Hamiltonian_SecularApprox}
\end{equation}

The Hamiltonian above is suitable to perform a conditional rotation, where the errors arising from the three approximations have been summed to provide a conservative estimate of the overall fidelity. The resulting gate rotates the spin by an angle $2\phi = \int{2\pi \gamma_e B_1(t)(\cos{\theta}+\sin{\theta})dt}$ within time $t$, described by the time evolution operator,
\begin{equation}
\mbox{$U(t)$}  = \left[
\begin{array}{cccc}
e^{(i\gamma_1)} & 0  & 0 & 0 \\
0 & e^{(i\gamma_2)} & 0 & 0 \\
0 & 0 & e^{(i\gamma_3)}\cos{\phi} & -i e^{(i\gamma_3)}\sin{\phi} \\
0 & 0 & -i e^{(i\gamma_3)}\sin{\phi}  & e^{(i\gamma_3)}\cos{\phi}
\end{array}
\right],
\label{eq:TimeEvoOp_t}
\end{equation}
where we have made the substitutions $\gamma_1 = -t(E_{\uparrow\uparrow}-(E_{\widetilde{\downarrow\uparrow}} - E_{\downarrow\downarrow}))$, $\gamma_2 = -t(E_{\widetilde{\uparrow\downarrow}})$ and $\gamma_3 = -t(E_{\widetilde{\downarrow\uparrow}})$ for compactness.  A single pulse of the above operator with $U(t:2\phi=\pi)$ yields an operation that closely resembles the CROT,
\begin{equation}
\mbox{$U_2$}  = \left[
\begin{array}{cccc}
e^{(i\gamma_1)} & 0  & 0 & 0 \\
0 & e^{(i\gamma_2)} & 0 & 0 \\
0 & 0 & 0 & -i e^{(i\gamma_3)} \\
0 & 0 & -i e^{(i\gamma_3)}  & 0
\end{array}
\right].
\label{eq:TimeEvoOp_tSimpleCROT}
\end{equation}

The above operation successfully rotates the spin of the second electron conditional upon the state of the first.  In order to complete the CROT (or CNOT) operation, however, the resulting phases must be accounted for.  The operator causes a phase shift for each of the four basis states $(\theta_{\uparrow\downarrow},\theta_{\downarrow\uparrow},\theta_{\uparrow\uparrow},\theta_{\downarrow\downarrow})$ which can be easily extracted by taking the phase of the non-zero element in the associated column for each basis state.  It is useful to analyze, instead, the phases associated with electron-1 ($\theta_1$) and electron-2 ($\theta_2$), the phase due to the interaction ($\theta_{12}$) and the global phase ($\theta_g$).  Only $\theta_1$, $\theta_2$ and $\theta_g$ can be corrected for with single-qubit rotations, requiring that our CROT have $\theta_{12} = 0$.  For the above operator, $\theta_{12} = \frac{1}{4}(\theta_{\uparrow\downarrow}+\theta_{\downarrow\uparrow}-\theta_{\uparrow\uparrow}-\theta_{\downarrow\downarrow}) = \frac{1}{4}(\gamma_2 - \gamma_1)$.  One possible solution is to use a refocusing pulse to correct for this phase as follows:
\begin{eqnarray}
\label{eq:TimeEvoSequence}
U_3 = X_{2} U_{\sqrt{CROT}} X_{2} U_{\sqrt{CROT}}\hspace{1mm},
\end{eqnarray}
where $U_{\sqrt{CROT}}$ is $U(t:2\phi=\pi/2)$ and $X_2$ unconditionally flips the spin of the second (target) electron.  For the phase to be refocused, $X_2$ must have $\theta_{12} = 0$.  The two transitions, $\nu_{\uparrow\updownarrow}$ and $\nu_{\downarrow\updownarrow}$ must both undergo a $\pi$-rotation in the same amount of time. This is satisfied by the operator
\begin{equation}
\mbox{$X_2$}  = \left[
\begin{array}{cccc}
0 & -i e^{(i\gamma_2)}  & 0 & 0 \\
-i e^{(i\gamma_2)} & 0 & 0 & 0 \\
0 & 0 & 0 & -i e^{(i\gamma_3)} \\
0 & 0 & -i e^{(i\gamma_3)}  & 0
\end{array}
\right],
\label{eq:TimeEvoOp_DesiredX}
\end{equation}
where $\theta_{12} = 0$.  This may be implemented as a two-tone pulse with amplitude adjusted for the $(\cos\theta+\sin\theta)$ difference in the transition matrix elements between the two pairs of states.  The result of Eq.~\ref{eq:TimeEvoSequence} yields the full operator for the CROT that successfully completes a conditional rotation and cancels out the phase due to the interaction.

We have attempted to quantify the major factors limiting the fidelity of creating a CROT gate using the two donor system.  However, in any realistic experiment additional non-idealities will arise which are not quantified here, such as inhomogeneous magnetic fields, pulsing errors and phase errors due to fluctuations in the energies of the Hamiltonian.  We have estimated the effect of fluctuations in the $^{29}$Si bath on the $\nu_{\uparrow\updownarrow}$, $\nu_{\downarrow\updownarrow}$, $\nu_{\updownarrow\uparrow}$ and $\nu_{\updownarrow\downarrow}$ transition probabilities. We can obtain an order-of-magnitude estimate of the associated phase decoherence error by considering the gate time, $T_\text{\tiny{CROT}}$, in comparison with the phase coherence time, $T_2$.  We choose a conservatively long $T_\text{\tiny{CROT}} \sim10^{-6}$~s from the high-fidelity region of Fig.~\ref{figure3}(d) and $T_2 \sim 10^{-1}$~s \cite{Muhonen2013}.  The resulting phase error is in the order of $10^{-5}$ and, therefore, $\sim 10\times$ smaller than the smallest error in Fig.~\ref{figure3}(d), indicating that it might not limit the fidelity in the experiment.

\section{\textit{ESR} Spectrum}
\label{fingerprint}
\begin{figure}[t!]
\includegraphics[width=1\columnwidth]{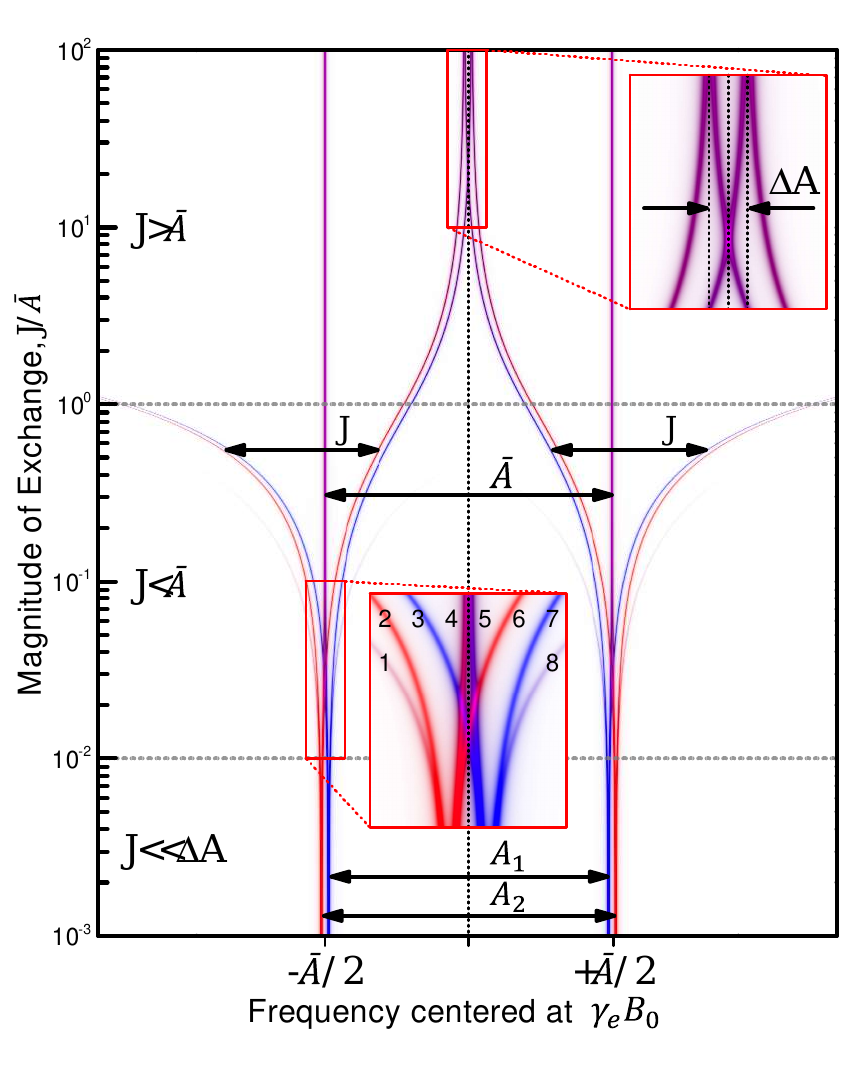}
\caption{\label{figure1} The ESR fingerprint of $H(J)$ plotted as a function of the exchange-hyperfine interaction ratio.  Branches are labeled from left to right (1 to 16) on the $J=10^{-1}\bar{A}$ line, with 1 to 8 shown.}
\end{figure}

The exchange coupling $J$ between the two donors in each interaction region needs to be calibrated in order to perform either of the proposed two-qubit logic gates.  Given a Hamiltonian $H(J)$, $J$ can be extracted from the ESR spectrum obtained by performing an experiment similar to that of Ref.~[\onlinecite{Pla2012}].  This `ESR fingerprint' [see Fig.~\ref{figure1}] is calculated by considering all transitions between the eigenstates of $H(J)$ in the regime $J < \gamma_e B_0$. Their respective intensities are weighted with (i) the electronic transition dipole matrix elements, $\langle{\psi_i|{(\sigma_{x_{S1}}+\sigma_{x_{S2}}})|\psi_f}\rangle$, where $\sigma_{x_{Sn}}$ is the Pauli operator for electron $n$, and (ii) the readout contrast, i.e. the change in expectation value of spin-$z$ projection of each electron upon excitation of the ESR transition, $\Delta \langle S_{n_z} \rangle = \langle\psi_f|S_{n_z}|\psi_f\rangle - \langle\psi_i|S_{n_z}|\psi_i\rangle$.  The transitions in Fig.~\ref{figure1} are color-coded such that the blue and red intensities are proportional to $\Delta \langle S_{1_z} \rangle$ and $\Delta \langle S_{2_z} \rangle$, respectively.  In the example of an allowed electronic transition, $|{\downarrow\downarrow\rangle}\rightarrow |{T_0\rangle}$, the line is drawn in purple as both $\Delta \langle S_{1_z} \rangle$ and $\Delta \langle S_{2_z} \rangle$ are non-zero. We plot in Fig.~\ref{figure1} an example where $A_2 > A_1$ and $\Delta A/\bar{A} = 2.5\%$.  The linewidth is taken to be smaller than $\Delta A$ such that this splitting can be resolved.


In the $J \ll \Delta A$ region, where the eigenstates are simply the combinations of logical electronic and nuclear states, we see a pair of blue and red lines that correspond to rotations of electron 1 and 2, respectively.  The transitions at $\gamma_e B_0 - A_i/2$ and $\gamma_e B_0 + A_i/2$ rotate electron $i$ when its binding nucleus is in the $|{\Downarrow\rangle}$ and $|{\Uparrow\rangle}$ state, respectively.  Thus, the red line at $\gamma_e B_0 - A_2/2$ includes the following four transitions: $|{\downarrow\downarrow\Downarrow\Downarrow\rangle} \rightarrow |{\downarrow\uparrow\Downarrow\Downarrow\rangle}$ (branch 1), $|{\uparrow\downarrow\Downarrow\Downarrow\rangle} \rightarrow |{\uparrow\uparrow\Downarrow\Downarrow\rangle}$ (branch 4), $|{\downarrow\downarrow\Uparrow\Downarrow\rangle} \rightarrow |{\downarrow\uparrow\Uparrow\Downarrow\rangle}$ (branch 2) and $|{\uparrow\downarrow\Uparrow\Downarrow\rangle} \rightarrow |{\uparrow\uparrow\Uparrow\Downarrow\rangle}$ (branch 6). In the following, we will focus our description of the transitions to the red branches on the left-hand side of the spectrum, noting that the same reasoning can directly be transferred to the other branches.

As $J$ becomes larger than the linewidth (region $J<\bar{A}$), we observe the exchange-splitting between transitions $|{\downarrow\downarrow\Downarrow\Downarrow\rangle} \rightarrow |{\downarrow\uparrow\Downarrow\Downarrow\rangle}$ (branch 1) and $|{\uparrow\downarrow\Downarrow\Downarrow\rangle} \rightarrow |{\uparrow\uparrow\Downarrow\Downarrow\rangle}$ (branch 4), and between transitions $|{\downarrow\downarrow\Uparrow\Downarrow\rangle} \rightarrow |{\downarrow\uparrow\Uparrow\Downarrow\rangle}$ (branch 2) and $|{\uparrow\downarrow\Uparrow\Downarrow\rangle} \rightarrow |{\uparrow\uparrow\Uparrow\Downarrow\rangle}$ (branch 6).
The electronic $|{\uparrow\downarrow\rangle}$ and $|{\downarrow\uparrow\rangle}$ states tend towards either the $|{T_0\rangle}$ and $|{S\rangle}$ states as $J/\Delta B_z$ increases for each particular nuclear configuration.  Branches 1 and 2 fade away for $J \sim \Delta A$ and $J \sim \bar{A}$, respectively, as they involve the state approaching a magnetically-inaccessible singlet state. Their $J$-split counterparts, branches 4 and 6, involving states approaching $|{T_0\rangle}$, tend towards $\gamma_eB_0 - \bar{A}/2$ and $\gamma_eB_0 - \Delta A/2$ (region $J>\bar{A}$), respectively.

A transition involving a fully entangled state would have $\Delta \langle S_{1_z} \rangle = \Delta \langle S_{2_z} \rangle =0.5$.  Accordingly, the region where the branches become purple indicate where a participating state tends towards the $|{T_0\rangle}$ or $|{S\rangle}$.

With the ability to independently prepare and readout the electron of each donor, it would be possible to observe every transition for a given $H(J)$.  The protocol would involve preparing or, at least, randomizing the nuclear spins using appropriate nuclear magnetic resonance (NMR) pulses, then extracting the ESR spectrum for both electrons as in Ref.~[\onlinecite{Pla2012}]. 
For a proof-of-principle device, we make the conservative assumption that only donor-2 is tunnel-coupled to a charge reservoir so that its electron spin can be read out in single-shot and initialized electrically in the ground state \cite{Morello2010}.  In this case, performing ESR experiments would only reveal those transitions involving $|{\downarrow\downarrow\rangle}$ with $\Delta \langle S_{2_z} \rangle > 0$.  In the $J \lesssim \Delta A$ regime, branches 1, 2, 11 and 13 could be observed.  In the $\Delta A \lesssim J \lesssim \bar{A}$ regime, branch 1 fades away and branch 5 emerges as its $\Delta \langle S_{2_z} \rangle$ increases.  Finally, in the $J \gtrsim \bar{A}$, branch 2 fades away while branch 10 emerges.  The relative spacing between the lines should make it easy to extract the value of $J$.  To be certain, slightly modifying $J$ by shifting the electrostatic environment will allow comparison to the fingerprint in Fig.~\ref{figure1}, ensuring correct interpretation of the observed resonances.  Operating in the $\gamma_e B_0 \gg \bar{A}$ regime, the resonant frequencies of these transitions are determined by four parameters ($\gamma_e B_0$, $\bar{A}$, $\Delta A$ and $J$) so that the determination of four resonant frequencies is sufficient to extract these parameters.

\section{Summary \& Conclusions}

We have analyzed the system of two exchange-coupled donor spin qubits, and shown how we can harness the hyperfine interaction with the $^{31}$P donor nuclei to implement two different types of two-qubit logic gates that form a universal set of operations when combined with single-qubit rotations. In the first method, we show that the amplitude of exchange oscillations can be controlled by exploiting the presence of the magnetic detuning, $\Delta B_z$, provided by the hyperfine interaction with the donor nuclei.  These oscillations can be switched on and off to form a $\sqrt{\text{SWAP}}$ gate of 99\% fidelity upon tuning $J$ by just two orders of magnitude.  In the second method, a two-qubit gate is implemented as the resonant rotation of one electron conditional upon the spin state of the other.  This method has the significant advantage that $J$ does not need to be tuned and a wide range of coupling strengths yields high-fidelity CROT gates in natural silicon ($>$95\%) and in isotopically-purified silicon ($>$99.99\%). Compared to previous proposals, our methods greatly relax the requirements on the accuracy of donor positioning and alignment of nanofabricated gates. We expect this will facilitate the construction of donor-based quantum information processors using ion implantation \cite{Persaud2005,Alves2013}, scanning-probe lithography \cite{Schofield2003}, or hybrid donor-dot devices \cite{Srinivasa2013}.


\begin{acknowledgments}
The authors thank F. A. Mohiyaddin, J. J. Pla, L. C. L. Hollenberg and A. S. Dzurak for useful discussions.  This research was funded by the Australian Research Council Centre of Excellence for Quantum Computation and Communication Technology (project number CE11E0096) and the US Army Research Office (W911NF-13-1-0024).
\end{acknowledgments}

\bibliography{Papers,RK_mar2013mendeley_AM}

\end{document}